\documentstyle[12pt,fleqn,epsf]{ioplppt}

\begin{document}

\title{Two and Three parametric regular generalizations of spherically
symmetric and axially symmetric metrics}

\author{N. N. Popov}

\address{Russian Academy of Sciences Computing Center,
         Vavilova St., 40,
         Moscow 117967, Russia}

\begin{abstract}
Regular generalizations  of  spherically and axially symmetric metrics
and   their    properties   are   considered.   Newton   gravity   law
generalizations are reduced for null geodesics.
\end{abstract}


\section{Introduction}

According to the  Birkgoff  theorem \cite{c01}  {\it  the metric of  a
vacuum spherically symmetric space time is  static,  singular  in  the
origin,  unique  and in  spherically  symmetric  coordinates  has  the
well-known Schwarzschild  form}  (when  $(+,-,-,-)$  signature form is
chosen)
\begin{equation}\label{e01}
ds^{2} = \biggl( 1 - \frac{a}{r}\biggr)  dt^{2}  -  \frac{dr^{2}}{1  -
\frac{a}{r}}   -   r^{2}   \biggl(d\theta   ^{2}   +    \sin^{2}\theta
d\varphi^{2}\biggr),
\end{equation}
where $a=2M$ is the Schwarzschild radius. We define  the regularity of
a metric at  a  point  as i) an indefinite  differentiability  of  all
metric  components  at this point; ii) the  condition  $det|g|\neq  0$
holds true at this point relative to any  coordinate system associated
with our system by a unique and continuous transformation.

Petrov  \cite{c02}  has  shown  a   counterexample   of   the   metric
contradicting the Birkgoff theorem
\begin{equation}\label{e02}
ds^{2} = \frac{t}{a  - t}  dt^{2} -  \frac{a  - t}{t}  dr^{2} -  t^{2}
\biggl(d\theta^{2} + \sin^{2} \theta d\varphi^{2} \biggr),
\end{equation}
where $t<a$. Metric (\ref{e02}) is not static. Here it is necessary to
note  that  (\ref{e01})  and  (\ref{e02})  are  defined  in  the  same
coordinate system. Further,  one can find another argument against the
uniqueness of  (\ref{e01}  ).The vacuum Einstein equations $R_{ij}=0$,
$i,j=0,...3$  are  nonlinear  second  order  differential   equations.
Therefore,  according   to   the   general   theory   of  differential
equations\cite{c03}  one  should  have  at  least   two  constants  of
integration in the solution, but Schwarzschild  metric (\ref{e01}) has
only one.

In this Letter we would like  to show that it is possible to construct
more  general  metrics  in  the  spherically   symmetric  and  axially
symmetric cases. Because of the Letter format we have to  restrict the
consideration only to the  main  results without proving the theorems,
which  we  plan  to  do in  the next  paper.  The  question  about the
applicability of our results in brane  models  with  noncompact  extra
dimensions and in astrophysics will  also  be  discussed in subsequent
publications.

\section{Schwarzschild case}

{\bf  Theorem  1. }{\it The} {\it spherically  symmetric,  static  and
asymptotically free metric, which is a  solution  of  vacuum  Einstein
equations $R_{ij}=0$, $i,j=0,...3$ has the following form:}
\begin{eqnarray}
ds^{2}  &  =  &  \Biggl(1 - \frac{a}{\sqrt[3]{r^{3}  +  b^{3}}}\Biggr)
dt^{2}   -   \frac{   r^{4}  dr^{2}}{\Biggl(\sqrt[3]{r^{3}  +   b^{3}}
\Biggr)^{4} \Biggl(1  -  \frac{a}{\sqrt[3  ]{r^{3}  + b^{3}}} \Biggr)}
\nonumber \\
& -  &  \biggl(\sqrt[3]{r^{3}  + b^{3}}\biggr)^{2}\biggl(d\theta^{2} +
\sin^{2} \theta d\varphi^{2} \biggr), \label{e03}
\end{eqnarray}
{\it  where $a$ is  the  Schwarzschild  radius  and $b\geq  0$  is  an
additional parameter  of the metric. This  metric is unique  to within
conformal transformations .}

{\bf  Conclusion  1.}  {\it  At $b=0$ we obtain  Schwarzschild  metric
(\ref{e01}).}

{\bf Conclusion 2.}  {\it At $b=a$  we obtain particle-  like  version
(analogously to  Ref.  \cite{c05}) of Schwarzschild metric \cite{c04}.
It  was  first   obtained   by  K.Schwarzschild  under  the  condition
$det|g_{ij}|=1$, which  has  violated  the  general  covariance of the
theory.}

Thus, metric (\ref{e03}) has the following properties:

\begin{enumerate}

\item
when $b=0$ (the classical Schwarzschild case)  it  has  the  curvature
singularity at $r=0$ and the coordinate singularity at $r=a$;

\item
when $0<b<a$ it has only  the  coordinate  singularity at $r=\sqrt[3]{
a^{3}-b^{3}}$;

\item
when $b\geq a$ metric (\ref{e03}) is regular everywhere.

\end{enumerate}

The  fulfillment  of  the  last  contention  can be  verified  by  the
transformation to the Kruskal-type coordinates
\[
v = e^{\frac{R  + t}{2a}}\sqrt{\frac{R}{a} - 1}, \qquad u=e^{\frac{R -
t}{2a}} \sqrt{ \frac{R}{a} - 1},
\]
where $R=\sqrt[3]{r^{3}+b^{3}}$. Accordingly, metric (\ref{e03}) takes
the form
\[
ds^{2} = -\frac{4a^{3}}{R}e^{ - \frac{R}{a}} du  dv  -  R^{2}  \biggl(
d\theta^{2} + \sin^{2} \theta d\varphi^{2}\biggr).
\]

It  is  necessary  to call the  reader's  attention  to  the fact that
formally (\ref{e03}) corresponds to all the metric requirements of the
differentiable manifold  only in the case  $b\geq a$. The  case $0\leq
b<a$ requires singular coordinate transformations,  and  that  is  not
completely correct from the formal  mathematical  point  of view while
working with smooth manifolds.

{\bf Remark.} {\it By chainging radial coordinate }${\it  r}${\it \ to
}$R$ {\it metric (\ref{e03})  may  be represented in the Schwarzschild
form}
\[
ds^{2} =  \biggl(1  -  \frac{a}{R}\biggr)dt^{2}  -  \frac{dR^{2}}{1  -
\frac{a}{R}}  -   R^{2}   \biggl(   d\theta^{2}   +   \sin^{2}  \theta
d\varphi^{2} \biggr).
\]

It seems that parameter $b$ can be cancelled completely, but  it would
not be correct. In fact parameter  $b$ is included into the $R$ domain
of  definition,  namely  $R\in  \lbrack b,\infty ).$ For  this  reason
variable $R$ is not  equal to the radial coordinate in all  the domain
of definition. Therefore, any attempt to  neglect  the  parameter  $b$
leads to a loss of  generality  of the results obtained. Moreover,  if
the Schwarzschild radius $  a=2M$ is defined in such a manner  that it
corresponds to Newton's  law  at $  r\rightarrow  \infty $, then,  the
value of $b$ only changes the geodesics deviations near the origin.

{\bf Lemma.} {\it  If one has  spherically symmetric space  time  with
metric ( \ref{e03}), the radial null geodesic equation is}
\[
\frac{d^{2}r}  {dt^{2}}  = -\frac{a}{2r^{2}}  \biggl(1  -  \frac{a}{R}
\biggr)  +   \frac{a\   (R-a)^{2}}  {Rr^{2}}  \Biggl(  \frac{1}{2R}  +
\frac{3}{2} \frac{1}{R - a} - 2 \frac{R^{2}}{ r^{3}} \Biggr).
\]
{\it At $r\rightarrow 0,$ the asymptotic behavior is:}
\begin{eqnarray}
\frac{d^{2}r}    {dt^{2}}    &    \rightarrow    &    -    \frac{3}{2}
\frac{a^{3}}{r^{4}} \qquad \mbox{when } b=0, \nonumber \label{e07} \\
\frac{d^{2}r}{dt^{2}}  &  \rightarrow  &  0 \qquad \mbox{when  }  b=a,
\nonumber \\
\frac{d^{2}r}{dt^{2}} & \rightarrow & - 2\frac{a\ b\ (b-a)^{2}}{r^{5}}
\qquad \mbox{when }b\neq a.
\end{eqnarray}

Theorem  1  perhaps improved  if  one could  cancel  the condition  of
staticity.  However,  the requirement of asymptotical flatness of  the
metric is essentially necessary due  to  the  Petrov example. Further,
the unexpected  behaviour of the  Newton gravity force near the origin
when  a particle  like  version of the  metric  (with the  appropriate
choice  of  the value  of  parameter $b$)  occurs.  The gravity  force
vanishes at the origin when $b=a$ and tends to infinity as $O(r^{-5})$
at  $b>a$.  Such  unusual  behaviour  may   be   responsible   for   a
gravitational   collapse   without   finishing   in   a   black   hole
configuration \cite{c06}.

\section{Other well-known metrics generalizations}

{\bf  Theorem  2.}   {\it   The  three  parametric  generalization  of
Reissner-Nordstrom metric is given by}
\[
ds^{2} = \Biggl(1 - \frac{a}{R} + \frac{Q^{2}}{R^{2}} \Biggr) dt^{2} -
\frac{r^{4}}{  R^{4}}  \frac{dr^{2}}{1 -  \frac{a}{R}  +  \frac{Q^{2}}
{R^{2}}}  -  R^{2}\biggl( d\theta ^{2} + \sin^{2} \theta  d\varphi^{2}
\biggr),
\]
{\it where $R=\sqrt[3]{r^{3}+b^{3}}$ and $Q$  is  a  U(1) type charge.
This metric is regular at  the  origin when $b>0$ and everywhere  when
$b\geq \frac{ a + \sqrt{a^{2} - 4 Q^{2}}}{2}$ and $a>2Q$.}

{\bf Theorem 3.} {\it The three parametric generalization  of the Kerr
metric is given by}
\begin{eqnarray}
ds^{2} &  = & \biggl(1  - \frac{aR}{R^{2} + \beta^{2} \cos^{2} \theta}
\biggr) dt^{2}  -  \frac{R^{2}  +  \beta^{2}  \cos^{2} \theta}{R^{2} +
\beta^{2} - aR} \frac{r^{4}}{ R^{4}} dr^{2}
- (R^{2} + \beta^{2} \cos^{2} \theta) d\theta^{2} \nonumber \\
&  -  &  \biggl(R^{2}  +  \beta^{2}  + \frac{\beta^{2}  a  R  \sin^{2}
\theta}{R^{2} +  \beta^{2}  \cos^{2}  \theta}  \biggr) \sin^{2} \theta
d\varphi^{2} +  \frac{ a \beta R  \sin^{2} \theta}{R^{2} +  \beta ^{2}
\cos^{2} \theta} dt d\varphi , \nonumber
\end{eqnarray}
{\it where $\beta  $  is  a specific inertia moment,  This  metric  is
regular everywhere when $b>0$ and  coordinate  singularities  are  not
taken into account.}

\section{Conclusions}

In this letter the metric that generalizes the well-known Schwarzchild
(Reissner-Nordstrom, Kerr) metric is considered . This metric includes
(as  a  partial  case)  the  Schwarzchild  metric (when an  additional
parameter $b$ vanishes), a class of metrics containing  only the event
horizon  without  the  origin  singularity  and,  finally, a class  of
singularity free  particle-like  metrics.  The  proof  of the theorems
follows. The applicability  of  this type  of  metric rather than  the
Schwarzchild  metric  to describe the existing astrophysical data  (on
accretion,  for  instance)  and  to  increase  the  generality  of our
knowledge of string black holes are under the investigation.

\section{Acknowledgments}

The author would like to thank S.Alexeyev and all the  participants of
the               Zelmanov               Memorial              Seminar
(http://xray.sai.msu.ru/sciwork/zelmanov) for the  useful  discussions
on the subject of this paper.

\begin{figure}[tbp]
\epsfxsize=0.5\hsize
\epsfysize=0.5\hsize
\centerline{\epsfbox{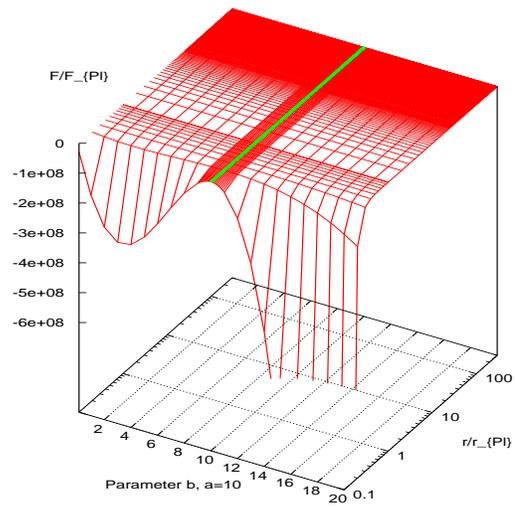}}
\caption{The  diagram  of  gravity  force dependency from  the  radial
coordinat $r$ and an additional parameter $b$ for null geodesic in the
vicinity of the origin in the\ case $a=10.$}
\end{figure}

\end{document}